\documentclass{article}

\usepackage{microtype}
\usepackage{graphicx}
\usepackage{subcaption}
\usepackage{booktabs} 

\usepackage{multirow}
\usepackage{multicol}
\usepackage{tabularx}

\usepackage{hyperref}

\usepackage[preprint]{icml2026}

\usepackage{amsmath}
\usepackage{amssymb}
\usepackage{mathtools}
\usepackage{amsthm}
\usepackage{bm}
\usepackage{empheq}
\usepackage{algorithm}
\usepackage{algpseudocode}
\usepackage{xurl}

\usepackage[capitalize,noabbrev]{cleveref}

\theoremstyle{plain}

\theoremstyle{definition}

\theoremstyle{remark}

\usepackage[textsize=tiny]{todonotes}

\icmltitlerunning{}

\newcommand{\methodname}{\textsc{MC$^2$Mark}}  

\begin{document}

\twocolumn[
  \icmltitle{\methodname: Distortion-Free Multi-Bit Watermarking for Long Messages}



  \icmlsetsymbol{equal}{*}

  \begin{icmlauthorlist}
    \icmlauthor{Xuehao Cui}{equal,yyy}
    \icmlauthor{Ruibo Chen}{equal,yyy}
    \icmlauthor{Yihan Wu}{yyy}
    \icmlauthor{Heng Huang}{yyy}
  \end{icmlauthorlist}

  \icmlaffiliation{yyy}{Department of Computer Science, University of Maryland, College Park, MD}

  \icmlcorrespondingauthor{Xuehao Cui}{cedrus@umd.edu}
  \icmlcorrespondingauthor{Ruibo Chen}{rbchen@umd.edu}
  \icmlcorrespondingauthor{Yihan Wu}{ywu42@umd.edu}
  \icmlcorrespondingauthor{Heng Huang}{heng@umd.edu}

  \icmlkeywords{Machine Learning, ICML}

  \vskip 0.3in
]



\printAffiliationsAndNotice{\icmlEqualContribution}

\begin{abstract}
Large language models now produce text indistinguishable from human writing, which increases the need for reliable provenance tracing. Multi-bit watermarking can embed identifiers into generated text, but existing methods struggle to keep both text quality and watermark strength while carrying long messages. We propose \methodname, a distortion-free multi-bit watermarking framework designed for reliable embedding and decoding of long messages. Our key technical idea is Multi-Channel Colored Reweighting, which encodes bits through structured token reweighting while keeping the token distribution unbiased, together with Multi-Layer Sequential Reweighting to strengthen the watermark signal and an evidence-accumulation detector for message recovery. Experiments show that \methodname\ improves detectability and robustness over prior multi-bit watermarking methods while preserving generation quality, achieving near-perfect accuracy for short messages and exceeding the second-best method by nearly 30\% for long messages.

\end{abstract}

\section{Introduction}

Recent advances in large language models have significantly improved text generation capabilities, but they also raise concerns about misuse and authenticity~\citep{goldstein2023generative, weidinger2021ethical}. To address these issues, statistical watermarking methods~\citep{Aaronson2022, kirchenbauer2023watermark, zhao2023provable, liu2023semantic} have been proposed to detect machine-generated text. These methods introduce hidden statistical patterns into the text generation process using secret keys, enabling later detection through statistical hypothesis testing. However, these methods often distort the model's output distribution, potentially degrading text quality.

To preserve generation quality while maintaining watermark effectiveness, distortion-free watermarking techniques have been developed. \citet{Aaronson2022} proposed Gumbel-max sampling to modify token distribution in a distortion-free manner. Subsequent approaches \citep{christ2023undetectable,kuditipudi2023robust,hu2023unbiased}, further optimized the trade-off between watermark strength and text quality. Recent methods like DiPmark \citep{wu2023dipmark}, STA-1 \citep{mao2024watermark}, and \textsc{MCmark} \citep{chen2025improved} have demonstrated improvements in robustness and detectability while maintaining distortion-freeness. However, these methods output only a zero-bit decision (watermarked or not) and do not provide the information needed for provenance tracing.

Multi-bit watermarking addresses this limitation by embedding additional information, such as model or user identifiers, into the generated text. Early approaches \citep{fernandez2023three, fairoze2023publicly} need access to the message space during detection, which limits their use in practice. More recent work, such as MPAC \citep{yoo2024advancing} and watermarks adopting error-correction codes~\citep{qu2025provably}, improved message encoding but sacrificed distortion-free properties. While BiMark \citep{feng2025bimark} achieved distortion-free multi-bit watermarking, its detectability diminishes with longer messages, posing challenges for real-world deployment. As pointed out by~\citet{feng2025bimark}, the problem to preserve text quality and achieve large message embedding capacity simultaneously is non-trivial because the two goals trade off.

We address this problem by proposing \methodname\, a distortion-free multi-bit watermark framework that supports reliable embedding and detection for long messages. \methodname\ uses Multi-Channel Colored Reweighting (MCCR), which dynamically scales token probabilities based on message bits while preserving the original distribution in expectation. We further propose Multi-Layer Sequential Reweighting (MLSR) to strengthen the statistical signals by iteratively reinforcing the watermark across layers. For message recovery, we use a detection method based on evidence accumulation to improve extraction accuracy.

The contribution of our work can be summarized as
\vspace{-2pt}
\begin{itemize}
\item We propose \methodname, a novel distortion-free watermarking framework for multi-bit message embedding and detection that maintains high detectability and robustness even for long messages.

\item We introduce Multi-Channel Colored Reweighting to embed messages by adaptively scaling token subsets, Multi-Layer Sequential Reweighting to strengthen watermark signals and a detection scheme based on evidence accumulation to robustly recover messages.

\item We empirically demonstrate that \methodname\ consistently outperforms existing muti-bit watermarking methods in detectablility and robustness, while maintaining generation quality. For 256-bit messages, \methodname\ achieves at least 91\% accuracy, which exceeds previous methods by nearly 30\%, setting a new benchmark for long-message multi-bit watermarking.
\end{itemize}

\section{Related Work}
\paragraph{Statistical watermarks.}
\citet{kirchenbauer2023watermark} extends the statistical watermarking framework first introduced by \citet{Aaronson2022} and validates its effectiveness on large language models.
Their method partitions the vocabulary into a red list and a green list, and biases generation toward green-list tokens.
To improve robustness, \citet{zhao2023provable} proposes a unigram watermark that uses one-gram hashing to construct watermark keys.
\citet{liu2023semantic} enhances robustness by integrating the semantic information of generated text into watermark keys.
In addition, \citet{liu2023unforgeable} uses neural networks to directly modify token probability distributions instead of predefined watermark keys, creating a strong watermarking scheme.
Despite their effectiveness, these methods can substantially alter the output distribution of the language model, which may degrade text quality.

\paragraph{Distortion-free watermarks.}
To preserve the quality of generated text, several studies have developed distortion-free watermarking methods.
\citet{Aaronson2022} uses prefix $n$-grams as watermark keys and applies the Gumbel-max trick to modify token distribution while preserving generation quality.
\citet{christ2023undetectable} determine watermark keys based on token positions and leverage inverse sampling during token sampling.
ITS-edit and EXP-edit \citep{kuditipudi2023robust} adjust the token distributions in a distortion-free manner using inverse sampling and Gumbel-max, respectively.
\citet{hu2023unbiased} introduce $\delta$-reweighting and $\gamma$-reweighting techniques, but requires prompts and logits from language models in detection.
DiPmark \cite{wu2023dipmark} further improves the $\gamma$-reweighting technique and introduces a robust detector.
\citet{mao2024watermark} show the tradeoff between quality and watermark strength in low-entropy generation settings and propose STA-1 for better quality and watermark strength.
\citet{dathathri2024scalable} introduced SynthID, which uses tournament-based watermarking to enable distortion-free generation.
Finally, \textsc{MCmark}~\citep{chen2025improved} splits token distribution into multiple channels and adjusts the distribution in a distortion-free manner, greatly improving detectability.

\paragraph{Multi-bit watermarks.}

Several studies have developed methods for embedding multi-bit messages in generated text.
\citet{fernandez2023three} assign each message a unique secret key to control the watermarking process. \citet{fairoze2023publicly} encodes messages by feeding them into hash functions that determine watermarking constraints during generation. However, these approaches need access to the message space or verification context during detection because the messages are bound to hash values or secret keys in ways that cannot be directly reversed from the watermarked text. 
\citet{yoo2024advancing} addresses this limitation by proposing Multi-bit watermark via Position Allocation (MPAC), which first assigns tokens to different parts of the messages and then encodes messages using the zero-bit watermarking scheme of \citet{kirchenbauer2023watermark}. \citet{qu2025provably} further improves robustness by incorporating error correction codes into multi-bit watermarking. However, these methods do not guarantee distortion-free generation.
BiMark~\citep{feng2025bimark} employs a layered bit-flip reweighting mechanism to achieve distortion-free generation, but its detectability degrades rapidly as the message length increases, which limits its application in real-world scenarios.
\begin{figure}[t]
    \centering
    \includegraphics[width=\linewidth]{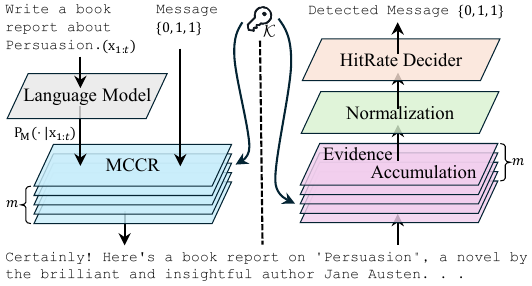}
    \caption{
    Overview of \methodname\ framework. The generation process (left) employs Multi-Channel Colored Reweighting (MCCR) and Multi-Layer Sequential Reweighting ($m$ layers illustrated) to produce text without distortion. The detection process (right) utilizes evidence accumulation for accurate message extraction.
    }
    \label{fig:workflow}
\end{figure}

\section{Method}
\begin{figure*}
    \centering
\includegraphics[width=0.99\linewidth]{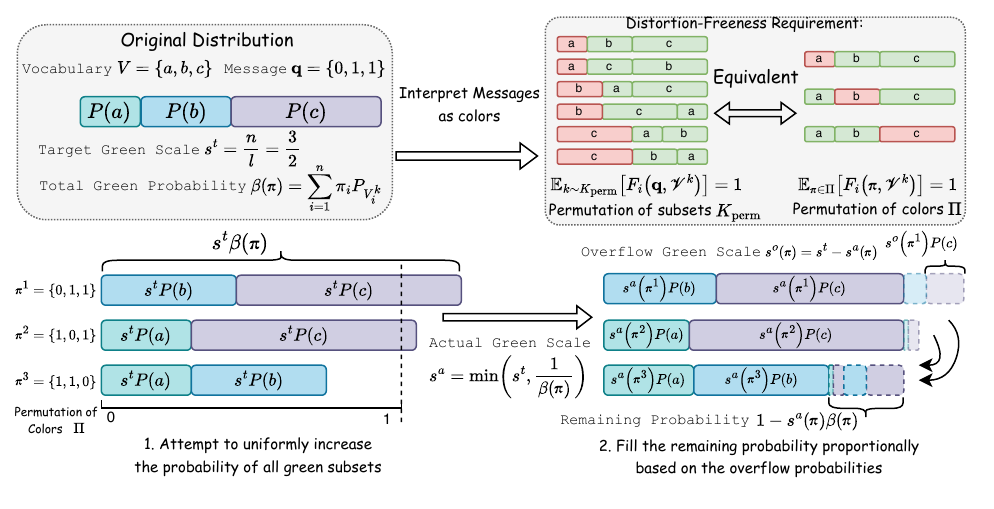}
    \caption{Illustration of the Multi-Channel Colored Reweighting (MCCR) framework. The proposed method interprets the message bit vector $\mathbf{q}$ as color assignments for partitioned vocabulary subsets $\mathcal{V}^k$, where $q_i=1$ designates a green subset and $q_i=0$ designates a red subset. The diagram visualizes the reweighting process across the permutation space $\Pi$ to maintain distortion-free generation. The algorithm first attempts to uniformly amplify green subsets via a target scale $s^t$, computes the actual feasible green scale $s^a(\boldsymbol{\pi})$ based on the total green probability $\beta(\boldsymbol{\pi})$, and subsequently redistributes the remaining probability mass using the overflow scale $s^o(\boldsymbol{\pi})$.}
    \label{fig:method}
\end{figure*}

\subsection{Notation}

We follow the notation established in prior studies~\citep{hu2023unbiased,wu2023dipmark,chen2025improved} to describe the token generation process of a large language model. Let $V$ denote the vocabulary of size $N = |V|$, and let $\mathcal{V}$ represent the set of all possible token sequences of arbitrary length, including the empty sequence. Given an input prompt, the model generates tokens autoregressively. We denote the probability of generating the next token $x_{t+1} \in V$, conditioned on the sequence of previously generated tokens $\mathbf{x}_{1:t} = (x_1, \ldots, x_t)$, as $P_M(x_{t+1} \mid \mathbf{x}_{1:t})$. The resulting conditional distribution over the vocabulary lies within $\mathcal{P}$, the probability simplex. For notational brevity, we omit the explicit dependency on the input prompt. Given a private key $k$ drawn from a key space $\mathcal{K}$, the reweighted logit distribution (referred to as a \textit{distribution channel} in~\citet{chen2025improved}) is denoted by $P_{M,w}(\cdot \mid \mathbf{x}_{1:t}, k)$. Existing literature~\citep{Aaronson2022,christ2023undetectable,hu2023unbiased,dathathri2024scalable} has emphasized the concept of \textit{distortion-free} (or unbiased) watermarks. This property ensures that, in expectation over the watermark key $k$, the token generation probabilities remain unaltered:
\begin{equation}
\mathbb{E}_{k\sim \mathcal{K}} \left[ P_{M,w}(x_{t+1} \mid \mathbf{x}_{1:t}, k) \right] = P_{M}(x_{t+1} \mid \mathbf{x}_{1:t}).
\end{equation}

Satisfying this condition implies that the quality of the generated text is better preserved.

\subsection{Multi-Channel Colored Reweighting}

We extend the Multi-Channel Watermark framework~\citep{chen2025improved} to encode multi-bit information, proposing the \textbf{Multi-Channel Colored Watermark} (\methodname). Given an $n$-bit message $\mathbf{c} \in \{0,1\}^n$ and a watermark key $k$, we first partition the vocabulary $V$ into $n$ disjoint subsets of equal size, denoted as $\mathcal{V}^k = \{V_1^k, V_2^k, \dots, V_n^k\}$. We then reweight the probability distribution of tokens within each subset using a scaling factor $\alpha_i^k \geq 0$. The watermarked distribution $P_{M,w}$ is defined as:
\begin{equation}
P_{M,w}(x_{t+1} \mid \mathbf{x}_{1:t}, k) = \alpha_i^k P_{M}(x_{t+1} \mid \mathbf{x}_{1:t}), \ \forall x_{t+1} \in V_i^k.
\end{equation}

As illustrated in Figure~\ref{fig:method}, we interpret the bit vector $\mathbf{c}$ as assigning a ``color" to each subset $V_i^k$. Specifically, if $c_i=1$, we designate $V_i^k$ as a \textbf{green} subset and aim to amplify its probability mass (i.e., $\alpha_i^k \ge 1$). Conversely, if $c_i=0$, we designate $V_i^k$ as a \textbf{red} subset and attenuate its mass (i.e., $\alpha_i^k \le 1$). To determine the optimal scaling factors $\boldsymbol{\alpha}^k \in \mathbb{R}^{n}$, we formulate an optimization problem where $\alpha_i^k$ is a function of the message and the partition:

\begin{equation} \label{eq:alpha}
\alpha^k_i = F_i(\mathbf{c}, \mathcal{V}^k).
\end{equation}

Our objective is to maximize the expected probability mass of the green subsets:\begin{equation}J = \mathbb{E}_{k \sim \mathcal{K}} \left[ \sum_{i=1}^n \alpha_i^k c_i P_{V_i^k} \right],\end{equation}where $P_{V_i^k} = \sum_{x \in V_i^k} P_M(x \mid \mathbf{x}_{1:t})$. This maximization is subject to the following constraints:


\begin{subequations}\label{eq:orig_constraint}
\begin{empheq}[left=\empheqlbrace]{align}
&\alpha_i^k \ge 0, \quad \forall i \in \{1,\dots,n\}, k \in \mathcal{K}, \label{eq:constraint-a} \\
&\sum_{i=1}^n \alpha_i^k P_{V_i^k} = 1, \quad \forall k \in \mathcal{K}, \label{eq:constraint-b} \\
&\begin{aligned}
    &\mathbb{E}_{k \sim \mathcal{K}} \left[ P_{M,w}(x_{t+1} \mid \mathbf{x}_{1:t}, k) \right] \\
    =& P_{M}(x_{t+1} \mid \mathbf{x}_{1:t}),
\end{aligned} \quad \forall x_{t+1} \in V. \label{eq:constraint-c}
\end{empheq}
\end{subequations}

Here, Eq.~\eqref{eq:constraint-b} ensures the watermarked distribution is valid (sums to 1), and Eq.~\eqref{eq:constraint-c} enforces the \textit{distortion-free} property. To satisfy the distortion-free constraint, we leverage the symmetry of the key space. For any $k \in \mathcal{K}$, there exists a set of keys $K_{\textrm{perm}} \subset \mathcal{K}$ such that the partition structure is identical, but the assignment of subsets is permuted. Specifically, for any $k' \in K_{\textrm{perm}}$, $P_{V_i^k} = P_{V_{\sigma(i)}^{k'}}$, where $\sigma$ is a permutation of $\{1, \dots, n\}$. Consequently, the distortion-free constraint (Eq.~\eqref{eq:constraint-c}) is satisfied if, for any such set $K_{\textrm{perm}}$, $\mathbb{E}_{k \sim K_{\textrm{perm}}} \left[ P_{M,w}(x_{t+1} \mid \mathbf{x}_{1:t}, k) \right] = P_{M}(x_{t+1} \mid \mathbf{x}_{1:t})$, i.e.,

\begin{equation}\label{eq:k_set}
\mathbb{E}_{k \sim K_{\textrm{perm}}} [F_i(\mathbf{c},\mathcal{V}^k)]=1,\forall i \in \{1,\dots,n\}
\end{equation}

Due to the symmetry of $\boldsymbol{\alpha}^k$ defined in Eq.~\eqref{eq:alpha}, permuting the subsets $\{V_i^k\}$ is equivalent to permuting the message bits $\mathbf{c}$. Let $l = \sum_{i=1}^n c_i$ be the Hamming weight of the message. We denote the set of all binary vectors with weight $l$ as:

\begin{equation}
\Pi = \left\{ \mathbf{x} \in \{0,1\}^n : \sum_{i=1}^n x_i = l \right\}.
\end{equation}

We can thus simplify Eq.~\eqref{eq:k_set} to the condition:

\begin{equation}\label{eq:pi_set}
\mathbb{E}_{\boldsymbol{\pi} \in \Pi} \left[ F_i(\boldsymbol{\pi}, \mathcal{V}^k) \right] = 1, \quad \forall i \in \{1, \dots, n\}.
\end{equation}

This transformation allows us to convert the original optimization problem into a linear programming task. However, finding the exact optimal solution is computationally prohibitive. We therefore propose a computationally efficient heuristic solution. As depicted in Figure~\ref{fig:method}, each $\boldsymbol{\pi} \in \Pi$ corresponds to a potential distribution channel. We must define the scaling factors for all $|\Pi| = \binom{n}{l}$ permutations such that Eq.~\eqref{eq:pi_set} holds.

First, we attempt to uniformly increase the probability of all green subsets. A subset $V_i^k$ is green if $\pi_i = 1$, which occurs in $\binom{n-1}{l-1}$ permutations. To satisfy Eq.~\eqref{eq:pi_set}, the ideal \textbf{target green scale} $s^t$ is:

\begin{equation}
s^t = \frac{\binom{n}{l}}{\binom{n-1}{l-1}} = \frac{n}{l}.\end{equation}

However, we must also respect the normalization constraint (Eq.~\eqref{eq:constraint-b}). Let $\beta(\boldsymbol{\pi}) = \sum_{i=1}^n \pi_i P_{V_i^k}$ represent the total probability mass of the green subsets for a given permutation $\boldsymbol{\pi}$. We define the \textbf{actual green scale} $s^a(\boldsymbol{\pi})$ as:

\begin{equation}
s^a(\boldsymbol{\pi}) = \min\left(s^t, \frac{1}{\beta(\boldsymbol{\pi})}\right).\end{equation}

The \textbf{overflow green scale} $s^o(\boldsymbol{\pi})$, which quantifies the scaling magnitude that cannot be accommodated by the channel, is given by:\begin{equation}s^o(\boldsymbol{\pi}) = s^t - s^a(\boldsymbol{\pi}).\end{equation}We aggregate this excess mass for each subset $V_i^k$ across all permutations to compute the total overflow probability $P_{V_i^k}^o$:\begin{equation}P_{V_i^k}^o = \sum_{\boldsymbol{\pi} \in \Pi} \pi_i s^o(\boldsymbol{\pi}) P_{V_i^k}.\end{equation}Finally, to fully utilize the probability space, any remaining probability mass $1 - s^a(\boldsymbol{\pi})\beta(\boldsymbol{\pi})$ is distributed proportionally based on the overflow probabilities. The resulting reweighting function $F_i$ is:

\begin{equation}F_i(\boldsymbol{\pi}, \mathcal{V}^k) = \pi_i s^a(\boldsymbol{\pi}) + \frac{(1 - s^a(\boldsymbol{\pi})\beta(\boldsymbol{\pi})) P_{V_i^k}^o}{\left(\sum_{j=1}^n P_{V_j^k}^o\right) P_{V_i^k}}.\end{equation}

We verify that this solution satisfies both Eq.~\eqref{eq:constraint-b} and Eq.~\eqref{eq:pi_set} in Appendix~\ref{appendix:distortion_free}. Substituting the specific message $\mathbf{c}$ into the reweighting function, the final scaling factor $\alpha_i^k$ is:\begin{equation}\label{eq:final_alpha}\alpha_i^k = F_i(\mathbf{c}, \mathcal{V}^k) = c_i s^a(\mathbf{c}) + \frac{(1 - s^a(\mathbf{c})\beta(\mathbf{c})) P_{V_i^k}^o}{\left(\sum_{j=1}^n P_{V_j^k}^o\right) P_{V_i^k}}.\end{equation}

\subsection{Generation}

The proposed \methodname\ generation process operates sequentially through a multi-layer manner. In this section, we detail the algorithm at a single time step $t$, given the base language model distribution $P_M(x_{t+1} \mid \mathbf{x}_{1:t})$ and the $n$-bit message payload $\mathbf{q} \in \{0,1\}^n$ intended for embedding.\paragraph{Complexity Reduction via Segmentation.}Directly computing the reweighting factors for \methodname\ using Eq.~\eqref{eq:final_alpha} requires evaluating $|\Pi| = \binom{n}{l}$ channels. This yields a computational complexity of $O\left(\binom{n}{l}\right)$, which scales to $O(2^n/\sqrt{n})$ when $l \approx n/2$, becoming prohibitively expensive for large $n$. To mitigate this computational cost, we employ a segmentation strategy. We partition the message $\mathbf{q}$ into $g$ disjoint segments $\mathbf{q}_1, \dots, \mathbf{q}_g$, where each segment $\mathbf{q}_i \in \{0,1\}^{n'}$ and $n' = n/g$. We assume $n$ is divisible by $g$ without loss of generality.

At each time step $t$, we derive a pseudo-random seed from the watermark key $k_t$ to sample: a segment index $\textrm{ind}_t \in \{1, \dots, g\}$, a masking vector $\mathbf{h}_t \in \{0,1\}^{n'}$ and a vocabulary partition $\mathcal{V}^{k_t} = (V^{k_t}_i)_{i=1}^{n'}$. The local payload for this step is computed as $\mathbf{q}'_t = \mathbf{q}_{\textrm{ind}_t} \oplus \mathbf{h}_t$, where $\oplus$ denotes the bit-wise XOR operation. This masking operation serves as a whitening transform, mitigating bias in $\mathbf{q}$ and ensuring that the expected Hamming weight satisfies $\mathbb{E}_{k \sim \mathcal{K}}[\sum_{i=1}^{n'} (\mathbf{q}'_t)_{i}] = n'/2$. 

\paragraph{Single-Layer Reweighting.} Subsequently, we compute the reweighting scales $\alpha_i^{k_t} = F_i(\mathbf{q}'_t, \mathcal{V}^{k_t})$ for all $i \in \{1, \dots, n'\}$ using Eq.~\eqref{eq:final_alpha} and update the distribution. For all $x_{t+1} \in V_i^{k_t}$, the watermarked probability is:
\begin{equation}\label{eq:single_layer}P_{M,w}(x_{t+1} \mid \mathbf{x}_{1:t}, k_t) = \alpha_i^{k_t} P_M(x_{t+1} \mid \mathbf{x}_{1:t}).
\end{equation}

\paragraph{Multi-Layer Sequential Reweighting.}We extend this approach to a multi-layer setting to enhance robustness, following the framework proposed by~\citep{dathathri2024scalable,wu2025ensemble,feng2025bimark}. We utilize $m$ distinct watermark keys $k_t^1, \dots, k_t^m$ at each step $t$. Let $P_0(\cdot \mid \mathbf{x}_{1:t}) = P_M(\cdot \mid \mathbf{x}_{1:t})$ denote the initial distribution. For each layer $j = 1, \dots, m$, we sequentially apply the \methodname\ reweighting. Specifically, given the partition defined by $k_t^j$, for any token $x_{t+1} \in V_i^{k_t^j}$, the distribution is updated as:
\begin{equation}
P_j(x_{t+1} \mid \mathbf{x}_{1:t}) = \alpha_i^{k_t^j} P_{j-1}(x_{t+1} \mid \mathbf{x}_{1:t}),
\end{equation}

where $\alpha_i^{k_t^j}$ is calculated based on the probability mass of the partition in the previous distribution $P_{j-1}$. The final watermarked distribution is given by:

\begin{equation}P
_{M,w}(\cdot \mid \mathbf{x}_{1:t}, \{k_t^j\}_{j=1}^m) = P_m(\cdot \mid \mathbf{x}_{1:t}).
\end{equation}

To ensure the watermark is recoverable, the keys $k_t^j$ are generated deterministically using a cryptographic hash of the preceding tokens $\mathbf{x}_{1:t}$, consistent with the settings in previous works~\citep{dathathri2024scalable,chen2025improved}. Furthermore, since the single-layer \methodname\ is distortion-free (satisfying Eq.~\eqref{eq:constraint-c}) and the keys for each layer are sampled i.i.d. from $\mathcal{K}$, the composite multi-layer reweighting preserves the distortion-free property, as demonstrated in~\citet{dathathri2024scalable,wu2025ensemble}. The complete generation procedure is summarized in Algorithm~\ref{alg:generator}.

\subsection{Detection}
\label{sec:detection}

Given only the generated sequence $x_{1:T}$, the detector aims to recover the embedded message $\hat{\mathbf{q}}\in\{0,1\}^n$ without access to the original payload. As summarized in Algorithm~2, detection mirrors the generation procedure: for each position $t$ and layer $j$, we deterministically reconstruct the same watermark components used during generation and aggregate statistical evidence across time and layers.

\paragraph{Evidence accumulation.}
For each next-token realization $x_{t+1}$ at time step $t$, we regenerate the layer-specific watermark key $k_t^j$, the segment index $\mathrm{ind}_t^j\in\{1,\dots,g\}$, the mask vector $\mathbf{h}_t^j\in\{0,1\}^{n'}$, and the corresponding vocabulary partition
$\mathcal{V}^{k_t^j}=\{V^{k_t^j}_1,\dots,V^{k_t^j}_{n'}\}$.
We then identify the unique active subset index $i^*$ such that $x_{t+1}\in V^{k_t^j}_{i^*}$.

Because \methodname\ increases the sampling probability of \emph{green} subsets, observing $x_{t+1}\in V^{k_t^j}_{i^*}$ provides evidence that subset $i^*$ was colored green at $(t,j)$. Recall that subset $i$ is green if and only if the corresponding local payload bit equals one:
\begin{equation}
\bigl(\mathbf{q}_{\mathrm{ind}_t^j}\oplus \mathbf{h}_t^j\bigr)_i = 1.
\end{equation}
Let $u^*=(\mathrm{ind}_t^j-1)n' + i^*$ denote the global bit index corresponding to segment $\mathrm{ind}_t^j$ and local coordinate $i^*$. Inverting the XOR in Eq.~(18) yields the implied hypothesis for $q_{u^*}$:
if $(\mathbf{h}_t^j)_{i^*}=0$, then green implies $q_{u^*}=1$; if $(\mathbf{h}_t^j)_{i^*}=1$, then green implies $q_{u^*}=0$.
We record this evidence in per-bit hit counters $\mathbf{c}^{\mathrm{hit}}_0,\mathbf{c}^{\mathrm{hit}}_1\in\mathbb{Z}_{\ge 0}^n$, incrementing $\mathbf{c}^{\mathrm{hit}}_v(u^*)$ for the corresponding implied value $v\in\{0,1\}$.

\paragraph{Normalization and decision rule.}
A direct hit count can be biased because different bits may be exposed under different mask configurations across $(t,j)$. To correct for this effect, we additionally maintain \emph{total-opportunity} counters $\mathbf{c}^{\mathrm{total}}_0,\mathbf{c}^{\mathrm{total}}_1\in\mathbb{Z}_{\ge 0}^n$.
For each reconstructed segment $\mathrm{ind}_t^j$, we update $\mathbf{c}^{\mathrm{total}}_v$ for \emph{all} local indices $i\in\{1,\dots,n'\}$: if $(\mathbf{h}_t^j)_i=0$, then a green observation at $i$ would support $q_u=1$, where $u=(\mathrm{ind}_t^j-1)n' + i$, and we increment $\mathbf{c}^\mathrm{total}_1$ by 1; if $(\mathbf{h}_t^j)_i=1$, it would support $q_u=0$, then we increment $\mathbf{c}^\mathrm{total}_0$ instead.
Thus, $\mathbf{c}^{\mathrm{total}}_v(u)$ counts the number of times the masking configuration at $(t,j)$ would have allowed an observation to contribute evidence in favor of hypothesis $q_u=v$.

After processing the full sequence, we compute an empirical hit rate for each bit $i\in\{1,\dots,n\}$ under both hypotheses:
\begin{equation}
\begin{aligned}
\mathrm{HitRate}_v(i)=\frac{\left(\mathbf{c}^{\mathrm{hit}}_v\right)_i}{\max\left(1,\left(\mathbf{c}^{\mathrm{total}}_v\right)_i\right)}, v\in\{0,1\}
\end{aligned}
\end{equation}
We decode each bit by comparing these normalized rates:
$\hat{q}_i=1$ if $\mathrm{HitRate}_1(i)>\mathrm{HitRate}_0(i)$, and $\hat{q}_i=0$ otherwise.
This normalization effectively marginalizes over randomness induced by the model distribution and the masking process, isolating the watermark signal while aggregating evidence across all time steps and layers.

\begin{table*}[t]
\centering
\small
\caption{Performance comparison across datasets for different message lengths $n$. [\textbf{KEY}: Best]}
\setlength{\tabcolsep}{6pt}
\begin{tabularx}{\textwidth}{
  c l *{7}{>{\centering\arraybackslash}X}
}
\toprule
$n$ & Method & \textit{book\_report} & \textit{mmw\_story} & \textit{fake\_news} & \textit{dolly\_cw} & \textit{longform\_qa} & \textit{finance\_qa} & \textit{c4\_subset} \\
\midrule
\multirow{4}{*}{16}
 & MPAC($\delta=$1.0) & 89.94\% & 96.48\% & 91.94\% & 79.63\% & 77.75\% & 78.06\% & 88.63\% \\
 & MPAC($\delta=$1.5) & 98.31\% & 99.61\% & 97.94\% & 90.81\% & 90.13\% & 92.22\% & 98.19\% \\
 & BiMark & 99.06\% & 99.61\% & 98.81\% & 94.38\% & 97.09\% & 97.13\% & 99.62\% \\
 & \methodname & \textbf{100.00\%} & \textbf{100.00\%} & \textbf{100.00\%} & \textbf{98.55\%} & \textbf{100.00\%} & \textbf{100.00\%} & \textbf{100.00\%} \\
\midrule
\multirow{4}{*}{32}
 & MPAC($\delta=$1.0) & 78.25\% & 86.39\% & 81.16\% & 75.56\% & 72.48\% & 72.86\% & 81.50\% \\
 & MPAC($\delta=$1.5) & 87.41\% & 97.43\% & 91.50\% & 81.38\% & 78.17\% & 80.92\% & 91.28\% \\
 & BiMark & 94.41\% & 97.52\% & 95.44\% & 93.50\% & 95.02\% & 91.41\% & 97.44\% \\
 & \methodname & \textbf{99.88\%} & \textbf{100.00\%} & \textbf{100.00\%} & \textbf{100.00\%} & \textbf{99.92\%} & \textbf{99.77\%} & \textbf{100.00\%} \\
\midrule
\multirow{4}{*}{64}
 & MPAC($\delta=$1.0) & 72.03\% & 79.00\% & 74.23\% & 66.80\% & 64.43\% & 66.27\% & 71.88\% \\
 & MPAC($\delta=$1.5) & 81.97\% & 88.62\% & 82.67\% & 73.00\% & 69.35\% & 69.96\% & 81.02\% \\
 & BiMark & 88.13\% & 91.67\% & 88.40\% & 87.76\% & 87.43\% & 86.08\% & 91.77\% \\
 & \methodname & \textbf{99.94\%} & \textbf{99.71\%} & \textbf{98.89\%} & \textbf{99.39\%} & \textbf{98.97\%} & \textbf{98.48\%} & \textbf{99.80\%} \\
\midrule
\multirow{4}{*}{128}
 & MPAC($\delta=$1.0) & 64.74\% & 70.42\% & 66.45\% & 61.62\% & 59.11\% & 59.96\% & 65.45\% \\
 & MPAC($\delta=$1.5) & 74.35\% & 80.14\% & 72.78\% & 68.35\% & 66.48\% & 66.85\% & 74.10\% \\
 & BiMark & 74.82\% & 79.35\% & 77.58\% & 74.99\% & 75.50\% & 72.52\% & 80.04\% \\
 & \methodname & \textbf{99.38\%} & \textbf{98.71\%} & \textbf{96.30\%} & \textbf{98.39\%} & \textbf{94.49\%} & \textbf{95.51\%} & \textbf{99.55\%} \\
 \midrule
\multirow{4}{*}{256}
 & MPAC(1.0) & 60.81\% & 63.83\% & 60.72\% & 57.82\% & 55.96\% & 56.48\% & 60.60\% \\
 & MPAC(1.5) & 65.81\% & 70.42\% & 66.75\% & 61.86\% & 60.86\% & 60.69\% & 67.25\% \\
 & BiMark & 57.27\% & 60.99\% & 58.74\% & 57.62\% & 56.75\% & 53.87\% & 60.57\% \\
 & \methodname
 & \bfseries 96.19\% & \bfseries 97.76\% & \bfseries 97.41\%
 & \bfseries 95.36\% & \bfseries 95.47\% & \bfseries 91.34\% & \bfseries 97.39\% \\
\midrule
\multirow{4}{*}{512}
 & MPAC(1.0) & 57.55\% & 59.37\% & 57.74\% & 56.38\% & 55.77\% & 55.64\% & 57.71\% \\
 & MPAC(1.5) & 60.95\% & 64.21\% & 60.88\% & 57.31\% & 55.62\% & 56.31\% & 60.80\% \\
 & BiMark & 38.19\% & 41.18\% & 39.79\% & 37.77\% & 36.68\% & 36.82\% & 40.62\% \\
 & \methodname
 & \bfseries 92.70\% & \bfseries 92.69\% & \bfseries 92.04\%
 & \bfseries 91.70\% & \bfseries 87.24\% & \bfseries 87.21\% & \bfseries 92.39\% \\
\bottomrule
\end{tabularx}
\vspace{-6pt}
\label{tab:message_length_comparison}
\end{table*}

\begin{table*}[t]
\centering
\small
\caption{Robustness under random token replacement and Dipper paraphrasing for various message lengths $n$.  [\textbf{KEY}: Best]}
\setlength{\tabcolsep}{6pt}
\begin{tabular*}{\textwidth}{c@{\extracolsep{\fill}}lccccc}
\toprule
\multirow{2}{*}{$n$} & \multirow{2}{2.1cm}{\textbf{Method}} &
\multicolumn{4}{c}{\textbf{Random Token Replacement}} &
\multirow{2}{*}{\textbf{Dipper Paraphrasing}} \\
\cmidrule(lr){3-6}
 &  & \textbf{10\%} & \textbf{20\%} & \textbf{30\%} & \textbf{50\%} &  \\
\midrule
\multirow{4}{*}{16}
 & MPAC($\delta=$1.0) & 84.81\% & 78.31\% & 73.19\% & 63.00\% & 60.06\% \\
 & MPAC($\delta=$1.5) & 96.38\% & 93.81\% & 87.13\% & 71.00\% & 66.19\% \\
 & BiMark & 98.09\% & 93.11\% & 85.78\% & 65.94\% & 51.56\% \\
 & \methodname & \textbf{100.00\%} & \textbf{100.00\%} & \textbf{100.00\%} & \textbf{90.44\%} & \textbf{72.40\%} \\
\midrule
\multirow{4}{*}{32}
 & MPAC($\delta=$1.0) & 75.78\% & 71.91\% & 67.16\% & 59.69\% & 59.28\% \\
 & MPAC($\delta=$1.5) & 85.84\% & 81.94\% & 75.78\% & 62.50\% & 58.72\% \\
 & BiMark & 92.93\% & 85.07\% & 75.66\% & 60.26\% & 45.00\% \\
 & \methodname & \textbf{100.00\%} & \textbf{99.63\%} & \textbf{97.97\%} & \textbf{80.28\%} & \textbf{67.59\%} \\
\midrule
\multirow{4}{*}{64}
 & MPAC($\delta=$1.0) & 68.67\% & 66.20\% & 62.03\% & 56.09\% & 55.23\% \\
 & MPAC($\delta=$1.5) & 76.88\% & 71.98\% & 66.50\% & 57.86\% & 56.59\% \\
 & BiMark & 83.89\% & 75.45\% & 67.28\% & 54.22\% & 59.32\% \\
 & \methodname & \textbf{99.09\%} & \textbf{96.97\%} & \textbf{91.52\%} & \textbf{70.84\%} & \textbf{62.51\%} \\
\midrule
\multirow{4}{*}{128}
 & MPAC($\delta=$1.0) & 62.55\% & 60.40\% & 57.00\% & 53.79\% & 53.66\% \\
 & MPAC($\delta=$1.5) & 70.41\% & 66.74\% & 63.34\% & 57.90\% & 55.34\% \\
 & BiMark & 72.62\% & 64.66\% & 58.38\% & 49.48\% & 47.06\% \\
 & \methodname & \textbf{97.85\%} & \textbf{93.39\%} & \textbf{85.34\%} & \textbf{65.43\%} & \textbf{58.61\%} \\
\midrule
\multirow{4}{*}{256}
 & MPAC(1.0) & 58.23\% & 57.23\% & 54.79\% & 52.73\% & 52.55\% \\
 & MPAC(1.5) & 63.94\% & 60.75\% & 58.67\% & 54.37\% & 53.92\% \\
 & BiMark    & 55.00\% & 50.42\% & 45.97\% & 41.01\% & 51.30\% \\
 & \methodname
 & \bfseries 92.35\% & \bfseries 84.61\% & \bfseries 75.52\%
 & \bfseries 60.84\% & \bfseries 54.90\% \\
\midrule
\multirow{4}{*}{512}
 & MPAC(1.0) & 56.51\% & 55.62\% & 54.37\% & 53.00\% & 52.57\% \\
 & MPAC(1.5) & 58.85\% & 57.18\% & 55.75\% & 52.92\% & 52.61\% \\
 & BiMark    & 37.21\% & 34.21\% & 31.99\% & 28.50\% & 44.78\% \\
 & \methodname
 & \bfseries 84.81\% & \bfseries 76.76\% & \bfseries 68.84\%
 & \bfseries 56.78\% & \bfseries 54.01\% \\
\bottomrule
\end{tabular*}
\vspace{-8pt}
\label{tab:robustness_random_token}
\end{table*}

\section{Experiments}
We implemented our pipeline in Python using the PyTorch framework and conducted experiments on four NVIDIA RTX6000ada GPUs. We evaluate our method in the aspects of detectability, robustness and distortion-freeness. 
The baselines are 
MPAC~\citep{yoo2024advancing} and BiMark~\citep{feng2025bimark}. We follow their original configurations in the experiments. The number of layers $m$ defaults to $10$ in following experiments.

\vspace{-2pt}
\paragraph{Datasets and Models.} The datasets used for detectability and robustness evaluation include three MMW datasets~\citep{piet2023mark} (i.e., \textit{book\_report}, \textit{mmw\_story}, and \textit{fake\_news}), \textit{dolly\_cw} dataset~\citep{DatabricksBlog2023DollyV2}, two WaterBench~\citep{tu2023waterbench} datasets (i.e., \textit{longform\_qa} and \textit{finance\_qa}) and a subset randomly selected from C4 dataset~\citep{raffel2020exploring}, denoted as \textit{c4\_subset} in following sections. For unbaisedness evaluation, we use MBart~\citep{liu2020multilingual} for machine translation and BART~\citep{lewis2019bart} for text summarization, following previous works~\citep{hu2023unbiased,wu2023dipmark}. The language models used for evaluation are Qwen2.5-3B-Instruct models~\citep{qwen2.5}.

\subsection{Detectability}

\begin{figure}[h]
    \centering
    \includegraphics[width=\linewidth]{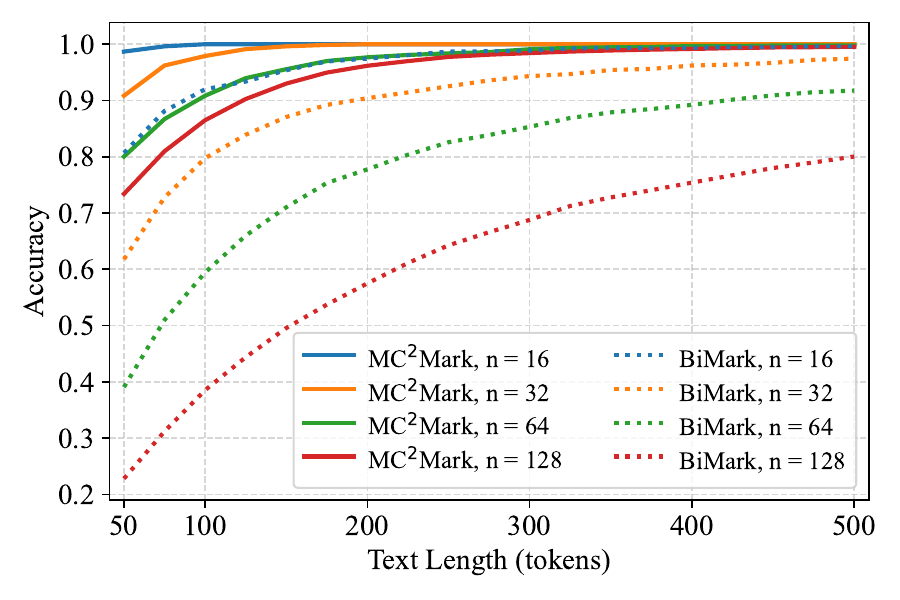}
    \caption{Detection accuracy as a function of text length for different message lengths $n \in \{16, 32, 64, 128\}$. Solid lines denote \methodname, while dotted lines denote BiMark.}
    \label{fig:detectability_curve}
    \vspace{-5pt}
\end{figure}

Table~\ref{tab:message_length_comparison} reports detection accuracy across seven datasets for different message lengths $n \in \{16, 32, 64, 128, 256, 512\}$. The comparison includes MPAC with $\delta=1.0$ and $\delta=1.5$, BiMark, and \methodname. The text length is 512 for all generation and evaluation. Across all datasets and message lengths, \methodname~consistently achieves the highest detection accuracy. When the message length is short, \methodname~reaches near-perfect or perfect accuracy on all datasets. As the message length increases, the performance of MPAC and BiMark degrades significantly across all datasets. In contrast, \methodname~maintains accuracy above 90\% in nearly all cases. When the message length is very long, \methodname~outperforms the second‑best method by roughly 30\%. These results show that \methodname~provides consistently strong and stable detectability across diverse domains and message lengths.

\vspace{-4pt}
\paragraph{Detectability vs. text length.} Figure~\ref{fig:detectability_curve} further analyzes detectability by illustrating detection accuracy as a function of text length for different message lengths. The comparison is conducted on \textit{c4\_subset} and focuses on \methodname\ and BiMark.
Across all message lengths and text lengths, \methodname\ exhibits stronger performance than BiMark. As the message length increases, the performance gap between the two methods becomes larger. For example, at $n=128$ and 50 tokens, BiMark achieves accuracy below 30\%, whereas \methodname\ remains above 70\%. Even with 500 tokens, BiMark fails to reach the accuracy level achieved by \methodname\ at much shorter text lengths. The accuracy of \methodname\ saturates quickly around 100\% as text length increases, suggesting that the embedded signal is evenly distributed across tokens and does not rely on long contexts for detection.

\begin{table*}[t]
\centering
\small
\caption{Impact of watermarking methods on text summarization and machine translation quality.}
\setlength{\tabcolsep}{6pt}
\begin{tabularx}{\textwidth}{
  l
  *{5}{>{\centering\arraybackslash}X}
}
\toprule
 & \multicolumn{3}{c}{\textbf{Text Summarization}} &
   \multicolumn{2}{c}{\textbf{Machine Translation}} \\
\cmidrule(lr){2-4} \cmidrule(lr){5-6}
\textbf{Method} & \textbf{BERTScore} & \textbf{ROUGE-1} & \textbf{Perplexity} &
\textbf{BERTScore} & \textbf{BLEU} \\
\midrule
No Watermark & 0.3058 & 0.3772 & 6.4155 & 0.5436 & 20.2038 \\
MPAC($\delta=$1.0) & 0.3045 & 0.3715 & 6.7869 & 0.5358 & 20.0009 \\
MPAC($\delta=$1.5) & 0.2897 & 0.3678 & 7.4193 & 0.5313 & 19.0335 \\
BiMark & 0.3172 & 0.3850 & 5.0777 & 0.5460 & 20.2671 \\
\methodname & 0.3065 & 0.3778 & 6.3742 & 0.5480 & 20.6516 \\
\bottomrule
\end{tabularx}
\label{tab:task_quality}
\vspace{-10pt}
\end{table*}

\subsection{Robustness}

Table~\ref{tab:robustness_random_token} reports the robustness of different methods under two types of perturbations: random token replacement and Dipper paraphrasing. For random token replacement, a fixed proportion of tokens (10\%, 20\%, 30\%, or 50\%) is randomly substituted. For paraphrasing, texts are rewritten using the Dipper model~\citep{krishna2023paraphrasing}. The evaluation is conducted for different message lengths $n \in \{16, 32, 64, 128, 256, 512\}$. Performance is measured by detection accuracy. Under the attack of random token replacement, \methodname\ demonstrates substantially stronger robustness across all message lengths and replacement ratios. For short messages ($n=16$), \methodname\ maintains perfect accuracy up to 30\% token replacement and remains above 90\% even when 50\% of tokens are replaced. Although performance gradually decreases with increasing message length, \methodname\ consistently outperforms all baselines in every setting. \methodname\ also achieves the highest robustness under Dipper paraphrasing attack for all message lengths.

\vspace{-3pt}
\subsection{Distortion-Freeness}

Table~\ref{tab:task_quality} evaluates whether different watermarking methods introduce systematic bias that degrades task performance. We compare text quality on two representative generation tasks: text summarization and machine translation. The ``No Watermark'' setting serves as the reference baseline.
For text summarization, our method maintains BERTScore, perplexity and ROUGE-1 values that are very close to the non-watermarked baseline, indicating minimal distortion of the summarization behavior. A similar pattern is observed for machine translation, in which our method achieves similar BERTScore and BLEU to the non-watermarked baseline.
The results demonstrate that the proposed watermarking approach is unbiased with respect to downstream task quality, preserving both semantic accuracy and fluency while embedding the watermark.

\subsection{Ablation Study}

\begin{figure}
    \centering
    \includegraphics[width=\linewidth]{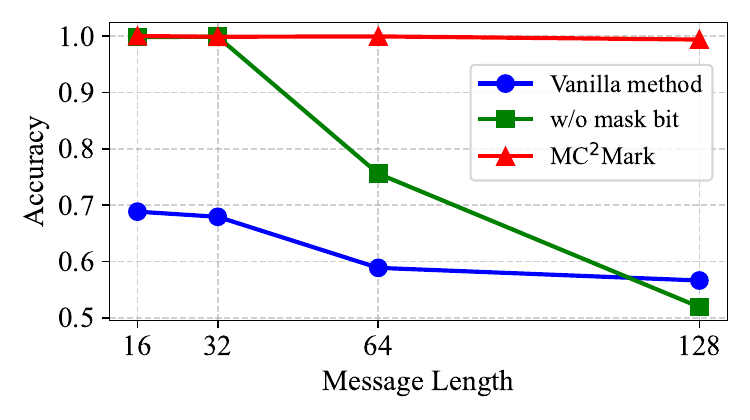}
    \caption{Accuracy comparison of MC$^2$Mark, MC$^2$Mark without the mask bit, and the vanilla method under different message lengths. The vanilla method directly applies \textsc{MCMark} which amplifies only one channel in reweighting.}
    \label{fig:ab1}
    
    \vspace{-4pt}
\end{figure}

\paragraph{Generation methods.} Figure~\ref{fig:ab1} reports an ablation study comparing different methods under the same experimental setting on \textit{book\_report} dataset. \methodname\ achieves the highest decoding accuracy for all message lengths, with only minor variation as the message length increases.
Removing the mask bit from \methodname\ leads to a noticeable performance drop when the message length becomes large. Although the accuracy at short message lengths ($n=16$ and $n=32$) remains close to that with the mask bit, the performance degrades significantly when $n=64$ and $n=128$. 
The vanilla method performs substantially worse than both \methodname\ variants.
Note that the detection method for \methodname\ does not apply to methods without the mask bit, in the case of which we use the tail bound of binomial distribution to detect the message.

\begin{figure}
    \centering
    \includegraphics[width=\linewidth]{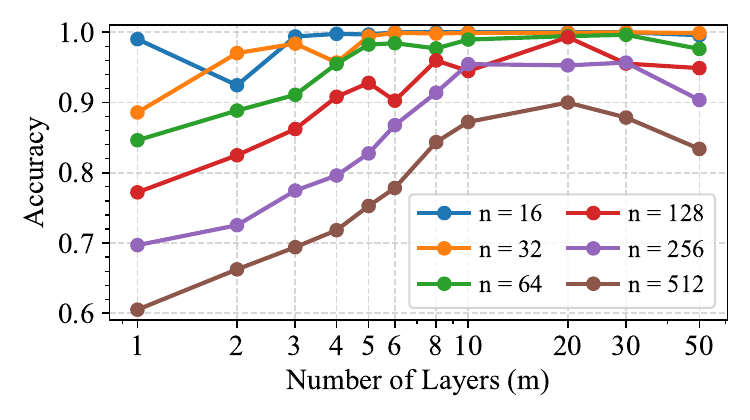}
    \caption{Accuracy as a function of the number of layers $m$ for different message lengths $n \in \{16, 32, 64, 128, 256, 512\}$ on \textit{longform\_qa} dataset. The horizontal axis is shown on a logarithmic scale. Performance tends to peak at $m=20$.}
    \label{fig:ab2}
\end{figure}

\vspace{-1pt}
\paragraph{Number of layers.} Figure~\ref{fig:ab2} presents an ablation study for the number of layers $m$ on \textit{longform\_qa} dataset. We evaluate \methodname\ with different number of layers $m$ while fixing all other settings, and report accuracy for different message lengths $n$. For short messages ($n=16$ and $n=32$), the accuracy is already high with a small number of layers. Increasing $m$ from 1 to around 10 layers leads to a rapid improvement, after which the performance saturates and remains close to 1.0. Adding more layers beyond this point does not yield noticeable gains and can slightly reduce accuracy in some cases, indicating diminishing returns. For longer messages, shallow \methodname\ perform worse, and accuracy increases more gradually as the number of layers grows. This shows that deeper architectures are required to capture the increased complexity associated with longer messages. For these settings, the performance tends to peak when the number of layers is approximately 20, and further increasing depth may incur slight loss of accuracy.

\section{Conclusion}
This work introduces \methodname, a distortion-free multi-bit watermarking framework that achieves large message embedding capacity while preserving text quality. Theoretical analysis and extensive experiments has shown that \methodname\ supports robust and accurate detection under long-message scenarios. The key lies in the layered multi-channel reweighting techniques and detection scheme base on evidence accumulation, providing a practical solution for deploying multi-bit watermarks in real world scenarios.

\newpage
\section{Impact Statement}
This paper presents \methodname, a distortion-free multi-bit watermarking framework that robustly embeds and detects long messages in generated text. The proposed method enables reliable provenance tracing of machine-generated content in public communication, which can serve as a technical foundation for the regulation of large language model–generated content and promote responsible use of large language models.

\bibliography{main}
\bibliographystyle{icml2026}

\newpage
\appendix
\onecolumn



\section{Distortion-Freeness}
\label{appendix:distortion_free}

First, we verify the satisfaction of Eq.~\eqref{eq:constraint-b}. Summing over all agents $i \in \{1, \dots, n\}$, we obtain:
\begin{align}
    \sum_{i=1}^n F_i(\boldsymbol{\pi},(V_i^k)_{i=1}^n)P_{V_i^k}
    &= s^a(\boldsymbol{\pi})\sum_{i=1}^n\boldsymbol{\pi}_i P_{V_i^k} + \sum_{i=1}^n\frac{(1-s^a(\boldsymbol{\pi})\beta(\boldsymbol{\pi}))P_{V_i^k}^o}{(\sum_{j=1}^n P_{V_j^k}^o)P_{V_i^k}}P_{V_i^k} \nonumber \\
    &= s^a(\boldsymbol{\pi})\beta(\boldsymbol{\pi}) + (1-s^a(\boldsymbol{\pi})\beta(\boldsymbol{\pi}))\frac{\sum_{i=1}^n P_{V_i^k}^o}{\sum_{j=1}^n P_{V_j^k}^o} \nonumber \\
    &= s^a(\boldsymbol{\pi})\beta(\boldsymbol{\pi}) + 1 - s^a(\boldsymbol{\pi})\beta(\boldsymbol{\pi}) = 1.
\end{align}

Next, we verify Eq.~\eqref{eq:pi_set}. We begin by expanding the expectation:
\begin{align}
    \mathbb{E}_{\boldsymbol{\pi} \in \Pi} \left[ F_i(\boldsymbol{\pi},(V_i^k)_{i=1}^n) \right]
    &= \frac{1}{|\Pi|}\sum_{\boldsymbol{\pi} \in \Pi} \left( \boldsymbol{\pi}_i s^a(\boldsymbol{\pi})+\frac{(1-s^a(\boldsymbol{\pi})\beta(\boldsymbol{\pi}))P_{V_i^k}^o}{(\sum_{j=1}^n P_{V_j^k}^o)P_{V_i^k}}\right) \nonumber \\
    &= \frac{1}{|\Pi|}\sum_{\boldsymbol{\pi} \in \Pi} \boldsymbol{\pi}_i s^a(\boldsymbol{\pi}) + \frac{1}{|\Pi|}\frac{P_{V_i^k}^o}{(\sum_{j=1}^n P_{V_j^k}^o)P_{V_i^k}}\sum_{\boldsymbol{\pi} \in \Pi}(1-s^a(\boldsymbol{\pi})\beta(\boldsymbol{\pi})). \label{eq:expansion_step}
\end{align}
To simplify the second term in Eq.~\eqref{eq:expansion_step}, we establish the following identity. Using the relationship $s^a(\boldsymbol{\pi}) = s^t - s^o(\boldsymbol{\pi})$ and the definition of $\beta(\boldsymbol{\pi})$, we derive:
\begin{align}
    \sum_{\boldsymbol{\pi} \in \Pi}(1-s^a(\boldsymbol{\pi})\beta(\boldsymbol{\pi}))
    &= \sum_{\boldsymbol{\pi} \in \Pi}(1 - s^t\beta(\boldsymbol{\pi}) + s^o(\boldsymbol{\pi})\beta(\boldsymbol{\pi})) \nonumber \\
    &= \sum_{\boldsymbol{\pi} \in \Pi}\left(1 - s^t\sum_{i=1}^n\boldsymbol{\pi}_i P_{V_i^k}\right) + \sum_{\boldsymbol{\pi} \in \Pi}\sum_{i=1}^n\boldsymbol{\pi}_i P_{V_i^k}s^o(\boldsymbol{\pi}) \nonumber \\
    &= |\Pi| - \frac{n}{l}\sum_{i=1}^n \left( P_{V_i^k} \sum_{\boldsymbol{\pi} \in \Pi}\boldsymbol{\pi}_i \right) + \sum_{i=1}^n P_{V_i^k}^o \nonumber \\
    &= |\Pi| - \frac{n}{l}\sum_{i=1}^n P_{V_i^k} \binom{n-1}{l-1} + \sum_{i=1}^n P_{V_i^k}^o.
\end{align}
Substituting $|\Pi| = \binom{n}{l} = \frac{n}{l}\binom{n-1}{l-1}$ and recalling that $\sum_{i=1}^n P_{V_i^k} = 1$, the expression simplifies to:
\begin{align}
    \sum_{\boldsymbol{\pi} \in \Pi}(1-s^a(\boldsymbol{\pi})\beta(\boldsymbol{\pi}))
    &= |\Pi| - |\Pi|\underbrace{\sum_{i=1}^n P_{V_i^k}}_{1} + \sum_{i=1}^n P_{V_i^k}^o = \sum_{i=1}^n P_{V_i^k}^o.
\end{align}
Finally, substituting this result back into Eq.~\eqref{eq:expansion_step} yields:
\begin{align}
    \mathbb{E}_{\boldsymbol{\pi} \in \Pi} \left[ F_i(\boldsymbol{\pi},(V_i^k)_{i=1}^n) \right]
    &= \frac{1}{|\Pi|}\sum_{\boldsymbol{\pi} \in \Pi} \boldsymbol{\pi}_i s^a(\boldsymbol{\pi}) + \frac{1}{|\Pi|}\frac{P_{V_i^k}^o}{(\sum_{j=1}^n P_{V_j^k}^o)P_{V_i^k}} \left( \sum_{j=1}^n P_{V_j^k}^o \right) \nonumber \\
    &= \frac{1}{|\Pi|}\sum_{\boldsymbol{\pi} \in \Pi} \boldsymbol{\pi}_i s^a(\boldsymbol{\pi}) + \frac{1}{|\Pi|}\sum_{\boldsymbol{\pi} \in \Pi} \boldsymbol{\pi}_i s^o(\boldsymbol{\pi}) \nonumber \\
    &= \frac{1}{|\Pi|}\sum_{\boldsymbol{\pi} \in \Pi} \boldsymbol{\pi}_i (s^a(\boldsymbol{\pi}) + s^o(\boldsymbol{\pi})) \nonumber \\
    &= \frac{s^t}{|\Pi|} \sum_{\boldsymbol{\pi} \in \Pi} \boldsymbol{\pi}_i
    = \frac{n/l}{\binom{n}{l}} \binom{n-1}{l-1} = 1.
\end{align}

\newpage

\section{\methodname\ Generator}
\begin{algorithm}[h]
\caption{\methodname\ Generation Algorithm}\label{alg:generator}
\begin{algorithmic}[1]
\Require Language model $M$, generation length $T$, vocabulary $V$, key space $\mathcal{K}$, message $\mathbf{q} \in \{0,1\}^n$, segments $g$, layers $m$.
\Ensure Watermarked sequence $\mathbf{x}_{1:T}$.
\State Calculate segment length $n' \leftarrow n/g$.
\For{$t=1$ \textbf{to} $T$}
    \State Initialize $P_0(\cdot \mid \mathbf{x}_{1:t}) \leftarrow P_M(\cdot \mid \mathbf{x}_{1:t})$ 
    \For{$j=1$ \textbf{to} $m$}
        \State Derive watermark key $k_t^j \in \mathcal{K}$ based on $\mathbf{x}_{1:t}$.
        \State Generate pseudorandom components using $k_t^j$: segment index $\textrm{ind}_{t}^j \in \{1, \dots, g\}$, mask vector $\mathbf{h}_t^j \in \{0,1\}^{n'}$, partition $\mathcal{V}^{k_t^j} = \{V_1^{k_t^j}, \dots, V_{n'}^{k_t^j}\}$
        \State Compute target message segment: $(\mathbf{q}')_t^j \leftarrow \mathbf{q}_{\textrm{ind}_t^j} \oplus \mathbf{h}_t^j$.
        \State Compute scaling factors $\alpha^{k_t^j}_i$ via Eq.~\eqref{eq:final_alpha} using $P_{j-1}$ statistics: $\alpha^{k_t^j}_i \leftarrow F_i((\mathbf{q}')_t^j, \mathcal{V}^{k_t^j}), \quad \forall i \in \{1, \dots, n'\}$
        \State Update distribution for all $i$ and all $x_{t+1} \in V_{i}^{k_t^j}$: $P_j(x_{t+1} \mid \mathbf{x}_{1:t}) \leftarrow \alpha_i^{k_t^j} P_{j-1}(x_{t+1} \mid \mathbf{x}_{1:t})$
    \EndFor
    \State Sample next token $x_{t+1} \sim P_m(\cdot \mid \mathbf{x}_{1:t})$
\EndFor
\State \textbf{return} $\mathbf{x}_{1:T}$
\end{algorithmic}
\end{algorithm}

\newpage
\section{\methodname\ Detector}

\begin{algorithm}[h]\caption{\methodname\ Detection Algorithm}\label{alg:detector}\begin{algorithmic}[1]\Require Generated sequence $\mathbf{x}_{1:T}$, Vocabulary $V$, message length $n$, segment length $n'$, layers $m$.\Ensure Detected message $\hat{\mathbf{q}} \in \{0,1\}^n$.\State Calculate number of segments $g \leftarrow n/n'$.\State Initialize hit counters $\mathbf{c}_0^{\text{hit}}, \mathbf{c}_1^{\text{hit}} \in \mathbb{Z}^n$ to $\mathbf{0}$.\State Initialize total occurrence counters $\mathbf{c}_0^{\text{total}}, \mathbf{c}_1^{\text{total}} \in \mathbb{Z}^n$ to $\mathbf{0}$.\For{$t=1 \text{ to } T-1$}\For{$j=1 \text{ to } m$}\State Derive watermark key $k_t^j$, mask $\mathbf{h}_t^j$, segment index $\text{ind}_t^j$, and partition $\mathcal{V}^{k_t^j}$.\State Identify the active partition index $i^*$ such that $x_{t+1} \in V_{i^*}^{k^j_t}$.
\State \textit{// Update Evidence (Hits)}
    \State Calculate global index $u^* \leftarrow (\text{ind}_t^j-1)n' + i^*$.
    \If{$(\mathbf{h}_t^j)_{i^*} = 1$}
        \State $(\mathbf{c}_0^{\text{hit}})_{u^*} \leftarrow (\mathbf{c}_0^{\text{hit}})_{u^*} + 1$ \Comment{Evidence supporting $q_{u^*} = 0$}
    \Else
        \State $(\mathbf{c}_1^{\text{hit}})_{u^*} \leftarrow (\mathbf{c}_1^{\text{hit}})_{u^*} + 1$ \Comment{Evidence supporting $q_{u^*} = 1$}
    \EndIf

    \State \textit{// Update Normalization Counts (Total Occurrences)}
    \For{$i=1 \text{ to } n'$}
        \State Calculate global index $u \leftarrow (\text{ind}_t^j-1)n' + i$.
        \If{$(\mathbf{h}_t^j)_{i} = 1$}
            \State $(\mathbf{c}_0^{\text{total}})_{u} \leftarrow (\mathbf{c}_0^{\text{total}})_{u} + 1$
        \Else
            \State $(\mathbf{c}_1^{\text{total}})_{u} \leftarrow (\mathbf{c}_1^{\text{total}})_{u} + 1$
        \EndIf
    \EndFor
\EndFor

\EndFor\State \textit{// Decode Message by comparing hit rates}\For{$i=1 \text{ to } n$}\State $\text{HitRate}_0 \leftarrow (\mathbf{c}_0^{\text{hit}})_i / \max(1, (\mathbf{c}_0^{\text{total}})_i)$\State $\text{HitRate}_1 \leftarrow (\mathbf{c}_1^{\text{hit}})_i / \max(1, (\mathbf{c}_1^{\text{total}})_i)$\If{$\text{HitRate}_1 > \text{HitRate}_0$}\State $\hat{q}_k \leftarrow 1$\Else\State $\hat{q}_k \leftarrow 0$\EndIf\EndFor\State \textbf{return} $\hat{\mathbf{q}}$\end{algorithmic}\end{algorithm}

\section{Additional Results}
We present more results across different datasets on different large language models. Table~\ref{tab:add1} shows the detectability results on Llama-3.2-3B-Instruct~\citep{dubey2024llama}, and Table~\ref{tab:add2} shows the detectability results on Phi-3.5-mini-instruct~\citep{abdin2024phi}. These results indicate that \methodname\ can work effectively on different language models.

\begin{table*}[t]
\centering
\small
\caption{Performance comparison on Llama-3.2-3B-Instruct across datasets for different message lengths $n$. [\textbf{KEY}: Best]}
\setlength{\tabcolsep}{6pt}
\begin{tabularx}{\textwidth}{
  c l *{7}{>{\centering\arraybackslash}X}
}
\toprule
$n$ & Method & \textit{book\_report} & \textit{mmw\_story} & \textit{fake\_news} & \textit{dolly\_cw} & \textit{longform\_qa} & \textit{finance\_qa} & \textit{c4\_subset} \\
\midrule
\multirow{4}{*}{16}
 & MPAC(1.0) & 77.63\% & 83.07\% & 83.81\% & 69.44\% & 72.38\% & 68.84\% & 80.75\% \\
 & MPAC(1.5) & 88.06\% & 93.29\% & 95.06\% & 73.38\% & 87.50\% & 75.88\% & 89.69\% \\
 & BiMark & 93.22\% & 96.74\% & 96.44\% & 93.75\% & 95.70\% & 96.79\% & 97.75\% \\
 & \methodname & \textbf{100.00\%} & \textbf{100.00\%} & \textbf{100.00\%} & \textbf{100.00\%} & \textbf{100.00\%} & \textbf{100.00\%} & \textbf{100.00\%} \\
\midrule
\multirow{4}{*}{32}
 & MPAC(1.0) & 69.56\% & 75.23\% & 71.97\% & 62.41\% & 66.41\% & 65.77\% & 74.44\% \\
 & MPAC(1.5) & 80.91\% & 87.73\% & 82.34\% & 69.25\% & 74.78\% & 72.28\% & 85.84\% \\
 & BiMark & 88.24\% & 90.03\% & 91.02\% & 83.78\% & 88.78\% & 91.78\% & 94.34\% \\
 & \methodname & \textbf{100.00\%} & \textbf{100.00\%} & \textbf{100.00\%} & \textbf{100.00\%} & \textbf{100.00\%} & \textbf{100.00\%} & \textbf{100.00\%} \\
\midrule
\multirow{4}{*}{64}
 & MPAC(1.0) & 67.47\% & 68.60\% & 67.95\% & 61.63\% & 63.88\% & 62.67\% & 70.00\% \\
 & MPAC(1.5) & 75.00\% & 80.16\% & 76.19\% & 64.86\% & 69.09\% & 66.14\% & 77.31\% \\
 & BiMark & 78.32\% & 80.67\% & 82.40\% & 75.75\% & 81.34\% & 82.88\% & 87.14\% \\
 & \methodname & \textbf{99.98\%} & \textbf{100.00\%} & \textbf{99.97\%} & \textbf{100.00\%} & \textbf{100.00\%} & \textbf{99.97\%} & \textbf{99.97\%} \\
\midrule
\multirow{4}{*}{128}
 & MPAC(1.0) & 60.53\% & 64.97\% & 60.51\% & 57.87\% & 58.08\% & 57.23\% & 62.57\% \\
 & MPAC(1.5) & 67.16\% & 71.13\% & 69.86\% & 62.14\% & 63.55\% & 62.85\% & 70.56\% \\
 & BiMark & 66.38\% & 68.86\% & 69.78\% & 63.36\% & 69.74\% & 69.55\% & 74.17\% \\
 & \methodname & \textbf{99.64\%} & \textbf{99.98\%} & \textbf{99.88\%} & \textbf{96.80\%} & \textbf{99.77\%} & \textbf{99.70\%} & \textbf{99.94\%} \\
\midrule
\multirow{4}{*}{256}
 & MPAC(1.0) & 57.55\% & 57.76\% & 57.95\% & 54.02\% & 54.66\% & 53.77\% & 58.27\% \\
 & MPAC(1.5) & 62.92\% & 65.10\% & 62.37\% & 58.29\% & 58.28\% & 59.04\% & 65.82\% \\
 & BiMark    & 48.91\% & 52.20\% & 51.59\% & 47.47\% & 51.12\% & 50.52\% & 54.73\% \\
 & \methodname
 & \bfseries 98.45\% & \bfseries 99.43\% & \bfseries 99.10\%
 & \bfseries 96.09\% & \bfseries 98.40\% & \bfseries 99.32\% & \bfseries 99.61\% \\
\midrule
\multirow{4}{*}{512}
 & MPAC(1.0) & 55.81\% & 56.56\% & 57.19\% & 53.89\% & 54.10\% & 54.33\% & 56.13\% \\
 & MPAC(1.5) & 59.35\% & 59.42\% & 58.82\% & 55.63\% & 55.40\% & 56.20\% & 60.24\% \\
 & BiMark    & 31.58\% & 34.02\% & 34.07\% & 30.11\% & 33.34\% & 31.26\% & 34.88\% \\
 & \methodname
 & \bfseries 95.61\% & \bfseries 95.28\% & \bfseries 96.06\%
 & \bfseries 94.84\% & \bfseries 94.69\% & \bfseries 95.65\% & \bfseries 95.77\% \\
\bottomrule
\end{tabularx}
\label{tab:add1}
\end{table*}

\begin{table*}[t]
\centering
\small
\caption{Performance comparison on Phi-3.5-mini-instruct across datasets for different message lengths $n$. [\textbf{KEY}: Best]}
\setlength{\tabcolsep}{6pt}
\begin{tabularx}{\textwidth}{
  c l *{7}{>{\centering\arraybackslash}X}
}
\toprule
$n$ & Method & \textit{book\_report} & \textit{mmw\_story} & \textit{fake\_news} & \textit{dolly\_cw} & \textit{longform\_qa} & \textit{finance\_qa} & \textit{c4\_subset} \\
\midrule
\multirow{4}{*}{16}
 & MPAC(1.0) & 78.88\% & 86.52\% & 82.63\% & 75.69\% & 83.69\% & 75.81\% & 85.50\% \\
 & MPAC(1.5) & 92.56\% & 97.66\% & 93.31\% & 88.00\% & 94.16\% & 89.25\% & 95.19\% \\
 & BiMark & 95.66\% & 98.82\% & 97.83\% & 96.00\% & 97.21\% & 95.90\% & 98.19\% \\
 & \methodname
 & \bfseries 100.00\% & \bfseries 100.00\% & \bfseries 100.00\%
 & \bfseries 93.49\% & \bfseries 99.88\% & \bfseries 100.00\% & \bfseries 100.00\% \\
\midrule
\multirow{4}{*}{32}
 & MPAC(1.0) & 68.75\% & 74.45\% & 70.66\% & 67.34\% & 68.44\% & 66.44\% & 74.31\% \\
 & MPAC(1.5) & 80.50\% & 86.17\% & 81.84\% & 76.06\% & 81.11\% & 76.47\% & 84.34\% \\
 & BiMark & 88.96\% & 94.24\% & 92.55\% & 90.15\% & 92.12\% & 88.41\% & 93.47\% \\
 & \methodname
 & \bfseries 100.00\% & \bfseries 99.97\% & \bfseries 100.00\%
 & \bfseries 99.27\% & \bfseries 98.91\% & \bfseries 99.97\% & \bfseries 100.00\% \\
\midrule
\multirow{4}{*}{64}
 & MPAC(1.0) & 66.78\% & 71.00\% & 68.20\% & 66.38\% & 67.00\% & 65.96\% & 70.00\% \\
 & MPAC(1.5) & 71.33\% & 77.20\% & 75.28\% & 68.28\% & 72.66\% & 69.38\% & 76.44\% \\
 & BiMark & 80.01\% & 85.66\% & 83.43\% & 79.89\% & 82.10\% & 78.78\% & 86.19\% \\
 & \methodname
 & \bfseries 98.98\% & \bfseries 97.95\% & \bfseries 99.89\%
 & \bfseries 98.57\% & \bfseries 99.33\% & \bfseries 99.59\% & \bfseries 99.03\% \\
\midrule
\multirow{4}{*}{128}
 & MPAC(1.0) & 57.38\% & 61.29\% & 58.67\% & 58.13\% & 59.57\% & 58.46\% & 60.82\% \\
 & MPAC(1.5) & 66.30\% & 71.17\% & 68.70\% & 64.21\% & 67.35\% & 65.08\% & 70.74\% \\
 & BiMark & 68.45\% & 73.92\% & 69.41\% & 67.65\% & 69.62\% & 66.99\% & 73.28\% \\
 & \methodname
 & \bfseries 97.57\% & \bfseries 99.58\% & \bfseries 95.77\%
 & \bfseries 94.78\% & \bfseries 98.36\% & \bfseries 98.54\% & \bfseries 98.80\% \\
\midrule
\multirow{4}{*}{256}
 & MPAC(1.0) & 57.29\% & 58.75\% & 57.64\% & 55.75\% & 57.66\% & 56.95\% & 58.34\% \\
 & MPAC(1.5) & 61.27\% & 63.59\% & 60.68\% & 58.70\% & 61.89\% & 60.34\% & 63.16\% \\
 & BiMark    & 52.33\% & 56.82\% & 54.16\% & 51.40\% & 52.87\% & 50.77\% & 55.28\% \\
 & \methodname
 & \bfseries 92.57\% & \bfseries 94.98\% & \bfseries 92.59\%
 & \bfseries 91.20\% & \bfseries 93.20\% & \bfseries 94.08\% & \bfseries 93.95\% \\
\midrule
\multirow{4}{*}{512}
 & MPAC(1.0) & 54.21\% & 56.55\% & 56.13\% & 54.83\% & 55.42\% & 54.90\% & 56.44\% \\
 & MPAC(1.5) & 57.49\% & 59.33\% & 56.66\% & 56.55\% & 56.98\% & 57.10\% & 58.42\% \\
 & BiMark    & 34.22\% & 37.44\% & 35.51\% & 33.89\% & 34.45\% & 33.07\% & 36.15\% \\
 & \methodname
 & \bfseries 86.69\% & \bfseries 88.95\% & \bfseries 86.20\%
 & \bfseries 86.67\% & \bfseries 87.37\% & \bfseries 86.84\% & \bfseries 90.16\% \\
\bottomrule
\end{tabularx}
\label{tab:add2}
\end{table*}

\end{document}